\documentclass[amssymb,prb,aps,amsmath]{revtex4}

\usepackage[dvips]{graphicx}

\begin{document}

\title{Electron  scattering due to dislocation wall strain field in GaN layers}
\author{S. Krasavin\footnote {Electronic mail: krasavin@theor.jinr.ru}}
\affiliation{Joint Institute for Nuclear Research, Bogoliubov
Laboratory of Theoretical Physics, 141980 Dubna, Moscow Region,
Russia}
\date{\today}
\begin{abstract}
The effect of edge-type dislocation wall strain field  on the Hall
mobility in $n$-type epitaxial GaN was theoretically investigated
through the deformation potential within the relaxation time
approximation. It was found that this channel of scattering can
play a considerable role in the low-temperature transport at the
certain set of the model parameters. The low temperature
experimental data were fitted by including this mechanism of
scattering along with ionized impurity and charge dislocation
ones.

\end{abstract}

\maketitle

{\bf I. INTRODUCTION}

 As is known GaN  films are under extensive examination for many
years because of their promising application for the construction
of short-wavelength light emitting devices\cite{pear,edgar}.
However, the performance of these devices is limited by defects,
both native and impurity types. Native defects, in particular, are
threading dislocations with high densities ($10^8-10^{11}$
cm$^{-2}$) which are result from the large lattice mismatch
between epilayer and substrate\cite{kapol,lester,hey}.
Dislocations, being charged objects, act as scattering centers
(core effect) for carriers affecting the transverse mobility in
films\cite{bou}. In most studies in the context of the GaN layers
this Coulomb scattering has been
considered\cite{weim,ng,look,gurusin}. At the same time, in
addition to the core scattering, dislocations can give
contribution to the resistivity through deformation and
piezoelectric potentials\cite{bou,pen}. In GaN layers these
potentials associate with the strain field of dislocation arrays
which form low-angle grain boundaries or separate dislocations in
the specific cases\cite{weim}. However, the carrier scattering due
to piezoelectric potential has been found as negligibly small
within the bulk of GaN\cite{shi}.

{\bf II. MODEL}

 In this paper we theoretically investigate the contribution to the Hall mobility in GaN layers
from a wall of dislocations of edge type. The scattering of
electrons by dislocation wall (DW) is treated in the framework of
the deformation potential approach. In this case, the perturbation
energy of electron can be written in the standard
form\cite{bard,dex}
\begin{equation}
\label{eq1} \delta U({\bf r}) =G\Delta ({\bf r}),
\end{equation}
where $\Delta ({\bf r})$ is the dilatation of lattice around a
finite dislocation array, $G$ is the deformation-potential
constant.


 Let the threading dislocation segments with coordinates ($0,h$)
 along the ($0,0,1$) axis (perpendicular to the interface/layer plane)
 form a dislocation wall of finite length
 $2L$. The dilatation at the point ${\bf r}\ge 2L$ around such defect, then,
 can be found based on the so-called disclination model of grain boundaries
 and dislocations for isotropic medium\cite{dewit,kleman}. It takes
the form\cite{kras1}
$$
\Delta ({\bf r})=\frac{(1-2\sigma )}{(1-\sigma)}\frac{b}{4\pi p
}\Bigl(\ln\frac{\sqrt{\rho^{2}_{+}+z^2}-z}{\sqrt{\rho^{2}_{+}+(h-z)^2}+(h-z)}-
$$
\begin{equation}
\ln\frac{\sqrt{\rho^{2}_{-}+z^2}-z}{\sqrt{\rho^{2}_{-}+(h-z)^2}+(h-z)}\Bigr),
\end{equation}
where $\rho ^{2} _{\pm}=(x\pm L)^2+y^2$, $b$ is the Burgers'
vector of dislocation in the wall, $p$ is the distance between
dislocations in the wall, $\sigma $ is the Poisson constant. It is
interesting to note that Eq.(2) is also the exact formula for the
dilatation around the high-angle grain boundary as well as the
disclination dipole.

 The square of the matrix element of electron scattering in momentum states ${\bf
 k}$ to states ${\bf k^{'}}$  with the perturbation
energy given by Eq.(2) deduced  in\cite{kras1} as
\begin{equation}
|\langle {\bf k}|U(r)|{\bf k^{'}}\rangle|^{2} = \frac{32\pi
^{2}A^2}{(q_{z}^{2}+q_{\perp}^{2})^2 V^2}\frac{\sin
^{2}(q_{z}h/2)}{q_{z}^{2}}(1-J_{0}(2q_{\perp}L)),
\end{equation}
where $A=\dfrac{Gb (1-2\sigma )}{4\pi p(1-\sigma )}$. As one can
see from Eq.(2) and Eq.(3), the scattering due to perturbation
$\delta U({\bf r})$ is three dimensional, therefore $({\bf k}-{\bf
k^{'}})_{z}=q_{z}=\sqrt{q^2-q_{\perp}^2}\not=0$ in contrast to the
case of an infinitely long dislocation line along the
$(0,0,1)$-axis. Omitting the details of further calculations, we
can come to the following equation for the relaxation time due to
the strain field of DW
$$
\tau_{wall}^{-1}(k)=\frac{32A^2L^2n_{def}\pi m^{*}}{\hbar ^3
k_{\perp}{^2}}\Bigl(\mathrm S\mathrm i
(hk_{\perp})+\frac{\cos(hk_{\perp})}{hk_{\perp}}-\frac{1}{hk_{\perp}}\Bigr)
$$
\begin{equation}
\Bigl(J_{0}^2(2Lk_{\perp})+J_{1}^2(2Lk_{\perp})-\frac{1}{2Lk_{\perp}}J_{0}(2Lk_{\perp})J_{1}(2Lk_{\perp})\Bigr),
\end{equation}
where $n_{def}$ is the arial density of DW which is inversely
proportional to the square of the mean distance between grain
boundaries, $m^{*}$ is the effective mass of carrier, $J_{n}(t)$
are the Bessel functions, Si(x) is the sine integral function,
$k_{\perp}=(k_{x},k_{y})$ is the normal to the disclination line
component of the wave vector, $\hbar $ is the reduced Planck's
constant. When a film is thick ($k_{\perp}h>>1$, bulk regime)
Eq.(4) can be simplified to
$$
\tau_{wall}^{-1}(k)=\frac{16A^2L^2n_{def}\pi ^{2} m^{*}}{\hbar ^3
k_{\perp}{^2}}
$$
\begin{equation}
\Bigl(J_{0}^2(2Lk_{\perp})+J_{1}^2(2Lk_{\perp})-\frac{1}{2Lk_{\perp}}J_{0}(2Lk_{\perp})J_{1}(2Lk_{\perp})\Bigr),
\end{equation}

 The relaxation time due to Coulomb scattering at
charged dislocation lines can be written in the well-known form
as\cite{podor}
\begin{equation}
\tau _{dis}(k)=\frac{\hbar ^3\epsilon ^{2}c^2}{N_{dis}e^4f^{2}m
^{*}}\frac{(1+4\lambda _{d}^2k_{\perp}^2)^{3/2}}{\lambda _{d}},
\end{equation}
where $N_{dis}$ is the dislocation density in the wall, $c$ is the
distance between acceptor centers along the dislocation line, $f$
is the occupation rate of the acceptor centers along the
dislocation, $\epsilon $ is the dielectric constant, $e$ is the
electronic charge, $\lambda _{d} = \sqrt{\dfrac{\epsilon
k_{B}T}{e^2n}}$ is the Debye screening length with electron
concentration $n=N_{D}^{+}-N_{A}^{-}-f(N_{dis}/c)$, and $k_{B}$ is
the Boltzmann's constant. To determine the filling factor $f$, the
procedure from\cite{gurusin} has been used.

Assuming the validity of the non-degenerate statistics, we can
evaluate the DW  contribution $\mu _{wall}$ to the total mobility
using the well-known formula
\begin{equation}
\mu (k) =\frac{e\hbar^2}{{m^{*}}^{2}k_{B}T}\frac{\int\tau (k)
k_{i}^2f_{0}d^3k}{\int f_{0}d^3k}=\frac{e\langle \tau
\rangle}{m^{*}},
\end{equation}
where $k_i=k_{x(y)}$ is the planar component of the wave vector,
and $f_{0}$ is the Boltzmann distribution function.

 {\bf III. NUMERICAL RESULTS}

We first analyzied the drift mobility contribution due to grain
boundary strain field together with other important mechanisms of
scattering at low temperatures (ionized impurities, charged
dislocation lines). The component of the scattering due to ionized
impurities has been taken in the form of the Brooks-Herring-like
formula obtained on the basis of the partial-wave phase-shift
method\cite{mey1}. This approach yields the correct results for
low $T$ and high $n$ where the Born approximation can be false. It
takes the form
\begin{equation}
\frac{1}{\tau _{ii}(k)}=N_{I}v\sigma ^{B}(k)H_{0},
\end{equation}
where $N_{I}$ is the ionized impurity concentration, $v$ is the
electron velocity, $\sigma ^{B}(k)$ is the Born cross section, and
$H_{0}$ is the factor of correction obtained from the phase-shift
calculations\cite{mey1}.
 We found that the Coulomb dislocation scattering is major at free carriers
concentration $n< 10^{17}$cm$^{-3}$ when $N_{dis}\approx 10^{8}$
cm$^{-2}$ ($T<100$K), and above this $n$ the impurity scattering
dominates. In our calculations dislocation core scattering always
dominates above $N_{dis}\approx 10^{8}$ cm$^{-2}$. The numerically
calculated $\mu_{wall}$ on the basis of the Eqs.(4),(6) is shown
in Fig.1 as a function of temperature for some selected model
parameters together with other contributions. The
deformation-potential constant $G$  has been taken equal to $4$
eV, that corresponds to the typical values for semiconductors.
From our analysis we found that $\mu _{wall}\sim T^{3/2}$ (unlike
the case of the separate dislocations where $\mu \sim T$), and
this contribution can be maximal for some chosen set of the
parameters when the concentration of the DW is sufficiently high
($n_{def}\simeq 10^{10}$cm$^{-2}$) (see Fig.1).


The low temperature part of the experimental data from\cite{kell}
for two GaN samples with $N_{dis}=4\times 10^8$ cm$^{-2}$ and
$2\times 10^{10}$ cm$^{-2}$  on the sapphire substrate has been
fitted based on the formula for the Hall mobility $\mu
_{H}=e\langle \tau ^2\rangle/m^{*}\langle \tau \rangle $. Here the
total relaxation time is given by
\begin{equation}
\langle \tau \rangle =\langle \frac{1}{\tau _{wall}^{-1}+\tau
_{dis}^{-1}+\tau _{ii}^{-1}}\rangle.
\end{equation}
 The results are presented in Fig.2. As shown,
there is  agreement with the experimental data when $30$K$
<T<100$K for both samples. We found that the result of the fit
essentially depends on the distance between dislocations in the
wall $p$, as compared to other parameters, and, hence on the angle
of misorientation between grains ($\theta=\arcsin(b/2p)$).

Our preliminary results show that this effect of the angle
variation on the Hall mobility can be noticeable even for a narrow
interval of $\theta$ between $1^{\circ}$ and $5^{\circ}$. The
second very sensitive parameter in our calculations is the
position of the dislocation acceptor level energy referred to the
conduction band edge energy.

As noted above, Eq.(2) describes both the dilatation around the
low-angle grain boundary and disclination dipole. The concept of
disclination dipole has been applied to obtain the deformations
around high-angle grain boundaries and linear defects of the
rotational type\cite{li1}. In this connection, the observed in
hexagonal GaN layers deformations associated with the 5/7 and 4/8
rings and high-angle grain boundaries (see, for example,
Refs.\cite{potin},\cite{bere}) can be considered in the framework
of the disclination dipole model. The second point which should be
noted concerns the interfacial misfit dislocations\cite{kim}.
Misfit dislocation scattering along with considered in this paper
can be given in the framework of the multi-layer model proposed
in\cite{look2}.

{\bf IV. CONCLUSIONS}\\
 In this article we have theoretically investigated the
possible role of the strain field associated with the dislocation
wall on the mobility in hexagonal GaN layers. It has been found
that this contribution to the total transverse mobility can be
noticeable at low temperatures for given above densities of such
defects and deformation constant typical for semiconductors. Our
calculations show the core scattering due to charged dislocation
lines is dominant mechanism when $N_{dis}>10^9$cm$^{-2}$. This
supports findings in previous publications devoted to this
material. At lower dislocation densities the ionized impurity
scattering and strain field scattering can dominate.


\newpage

\pagestyle{empty}

 \centerline{Figure Captions} \vskip 1cm

\noindent Fig.1. Calculated contributions to the total drift
mobility as a function of temperature for the model parameters:
$N_{dis}=10^{8}$cm$^{-2}$, $n_{def}=1.5\times 10^{10}$ cm$^{-2}$,
$f=0.8$, n=$4\times 10^{16}$ cm$^{-2}$, $p=10^{-2}\mu$m,
$h=1.2\mu$m. The rest of the parameter set has been taken from
Ref.\cite{shur}

\vspace{1.5cm}

\noindent Fig.2. Hall mobility vs temperature for two samples with
$N_{dis}=4\times 10^{8}$cm$^{-2}$ (squares), $N_{dis}=2\times
10^{10}$cm$^{-2}$ (circles) from Ref. Solids lines are theoretical
curves. The set of the model parameters: (for squares) $G=9$ eV,
$f=0.92$, $p=9\times 10^{-3} \mu$m, $n_{def}=2.7\times 10^{10}$
cm$^{-2}$, (for circles) $G=7$ eV, $f=0.98$, $p=6\times 10^{-3}
\mu$m, $n_{def}=1.5\times 10^{10}$ cm$^{-2}$.

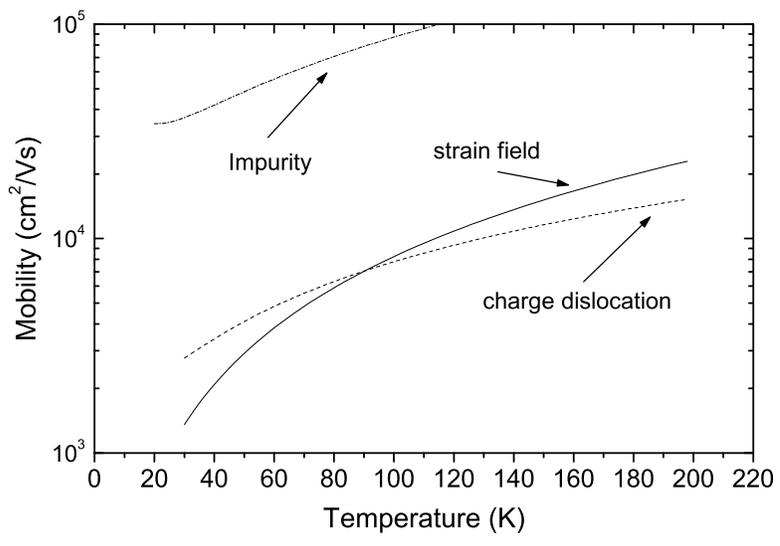
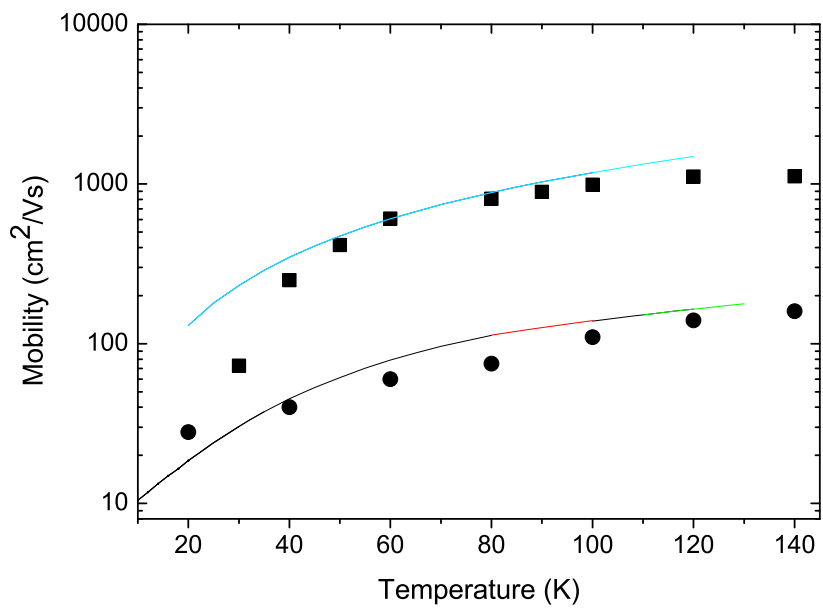
\end{document}